\documentclass[twocolumn,tightenlines,showpacs,prd,floatfix,preprintnumbers,amsmath,amssymb,nofootinbib,superscriptaddress]{revtex4}

\usepackage[dvips]{graphicx}
\usepackage{subfigure}

\newcommand{\F}{$\mathcal F$}

\begin{document}
\title{Implementation of barycentric resampling for continuous wave 
searches in gravitational wave data}
\author{Pinkesh Patel}
\affiliation{California Institute of Technology}
\email[]{ppatel@ligo.caltech.edu}
\author{Xavier Siemens}
\affiliation{University of Wisconsin Milwaukee}
\email[]{siemens@gravity.phys.uwm.edu}
\author{Rejean Dupuis}
\affiliation{}
\email[]{rejean@gmail.com}
\author{Joseph Betzwieser}
\affiliation{California Institute of Technology}
\email[]{josephb@ligo.caltech.edu}

\date{\today}

\begin{abstract}
We describe an efficient implementation of a coherent statistic for
continuous gravitational wave searches from neutron stars. The algorithm works by transforming the data taken by a gravitational wave detector from a moving Earth bound frame to one that sits at the Solar System barycenter. Many practical difficulties arise in the implementation of this algorithm, some of which have not been discussed previously.  These difficulties include constraints of small computer memory, discreteness of the data, losses due to interpolation and gaps in real data. This implementation is considerably more efficient than previous implementations of these kinds of searches on Laser Interferometer Gravitational Wave (LIGO) detector data.
\end{abstract}

\maketitle

\section{Introduction}

Rapidly rotating neutron stars are among the most promising sources of
continuous gravitational waves. They can emit gravitational waves
through a variety of mechanisms, including unstable oscillation
modes~\cite{Bildsten:1998ey, Andersson:1998qs} and deformations of the
crust~\cite{Bildsten:1998ey, Ushomirsky:2000ax, Cutler:2002nw,
Melatos:2005ez, Owen:2005fn}. Neutron stars can radiate powerful
beams of radio waves from their magnetic poles. If a neutron star's
magnetic poles are not aligned with its rotational axis, the beams
sweep through space, and if the Earth lies within the sweep of the
beams, the star is observed as a point source in space emitting bursts
of radio waves. Such a neutron star is called a pulsar~\cite{Gold:1968zf,1938}. Since the
first discovery~\cite{Hewish:1968bj}, around 2000 pulsars have been
detected~\cite{ATNF, Manchester:2004bp, lorimerlr}.

Due to magnetic dipole radiation and gravitational radiation, the rotational
frequencies of neutron stars slowly decrease in time.  Other than this
effect, gravitational waves from isolated rotating neutron stars are
essentially monochromatic in the rest frame of the star. The waves are
continuous and their frequency is determined by the rotational
frequency of the star. The motion of the detector as the earth rotates
about its axis and around the sun, however, modulates the phase as
well as the amplitude of the received signal. In order to recover the
signal from interferometric data optimally, both of these effects must
be taken into account. Detecting gravitational waves from neutron
stars could reveal information about the strength of neutron star
crusts and the equation of state of the nuclear matter that makes up
the star~\cite{Owen:2005fn}. Continuous gravitational waves may also
be produced by other sources, such as cosmic
strings~\cite{Dubath:2007wu,DePies:2009mf}.

There are a number of techniques available for continuous wave searches. These
techniques can be loosely divided into two categories: (1) coherent
methods~\cite{Jaranowski:1998qm, Abbott:2003yq}, which keep track of
the phase of the gravitational wave signals over long periods of time,
and (2) semi-coherent methods~\cite{Abbott:2007td}, which combine
shorter periods of data without tracking the phase (for example, taking
Fourier transforms of short segments of data and then summing the
power).

When the sky location and phase evolution of a neutron star are known,
a coherent search for continuous gravitational waves is relatively
straightforward~\cite{Abbott:2003yq}. Assuming that the noise in a gravitational wave detector follows Gaussian statistics, in the presence of a signal, the
signal to noise recovered in a search increases with the square root
of the amount of data used in the search. This is because the signal
amplitude grows linearly while the noise follows a random walk. Thus,
with enough data, it is possible to recover any continuous signal out of 
noisy data.

If certain parameters of the signal (sky location, frequency, spindowns and binary parameters) are not known the search becomes much more involved.  The reason is
that the number of points needed to cover the search parameter space
(and ensure no signals are lost) grows like a large power of the
amount of data used~\cite{Brady:1997ji}. This makes the sensitivity of gravitational wave searches computationally bound: One cannot simply integrate
arbitrary amounts of data to gain sensitivity because there is not
enough computational power available to perform the search. Thus, more
efficient code and greater computing power are highly desirable, since
they translate into more data being analyzed and therefore an increase
in the sensitivity of gravitational wave searches.

A promising method for blind searches involves exploiting large-scale correlations in the coherent detection statistic ~\cite{Pletsch:2009prl}. Another method that has been successful in these kinds of searches is the hierarchical scheme of incoherently combining coherent sets of data. Some of the methods currently under use include the Hough transform and stack-slide~\cite{Abbott:2007td}.

In this paper we focus on an efficient implementation of coherent
techniques.  The method we present here is similar to several previous implementations~\cite{Astone:2002prd,Astone:2003cqg,Krolak:2004prd}, in that it uses fast Fourier transforms (FFTs) to calculate the so-called ${\cal F}$-statistic~\cite{Jaranowski:1998qm}, the logarithm of the
likelihood function maximized over the intrinsic (and unknown)
parameters of the gravitational wave produced by a neutron star, but with one very important difference.  We resample the time domain data to the Solar System barycenter before taking a FFT.  This allows us to use a single FFT to calculate the detection statistic for arbitrarily many frequencies and an arbitrary amount of observation time, while previous implementations have a maximum frequency band and observation time that can be calculated with a single FFT, which are determined by losses due to phase mismatch.  Another set of techniques, described in~\cite{Torre,Braccini,Astone:2008cqg}, implement stroboscopic resampling methods described in~\cite{Schutz:Proceedings}. The stroboscopic method requires data at full bandwidth, and is therefore not suitable for distributed computing applications such as Einstein@Home.

In Section~\ref{sec:signal} we review the signal properties and the
nearly-optimal coherent statistic that can be used to extract continuous
signals from interferometric gravitational wave data. In Section~\ref{sec:implement_bary_resamp} we discuss how to implement the calculation of the nearly-optimal coherent statistic in a computationally efficient way in both the time and frequency domains.  In Section~\ref{sec:results} we describe the results of a computer code using this algorithm on software injections of gravitational waves into gaussian noise.  Lastly, in Section~\ref{sec:practical}
we address important technical issues to do with practical
implementations of barycentric resampling.

\section{Preliminaries}
\label{sec:signal}
In this section we closely follow the method of Jaranowski, Krolak, and
Schutz \cite{Jaranowski:1998qm} to provide the background on the
signal and the detection statistic. Power-recycled Fabri-Perot
Michelson interferometers such as those used by the Laser
Interferometer Gravitational Wave Observatory (LIGO) are sensitive to
the strain caused by gravitational waves passing through it.  The
strain measured at a detector can be written
as~\cite{Jaranowski:1998qm}
\begin{equation}
h(t) = F_{+}(t) h_{+}(t) + F_{\times}(t) h_{\times}(t),
\label{eq1}
\end{equation}
where $t$ is the time in the detector frame, and $h_{+}$ and
$h_{\times}$ are the ``plus'' and ``cross'' polarizations of gravitational
wave. $F_{+}(t)$ and $F_{\times}(t)$ are the beam-pattern
functions of the interferometer and are given by
\begin{equation}
F_{+}(t) = \sin{\zeta}[a(t)\cos{2\psi}+b(t)\sin{2\psi}],
\label{eq2}
\end{equation}
and
\begin{equation}
F_{\times}(t) = \sin{\zeta}[b(t)\cos{2\psi}-a(t)\sin{2\psi}],
\label{eq3}
\end{equation}
where $\psi$ is the polarization angle of the wave and $\zeta$ is the
angle between detector arms (which in the case of LIGO is 90$^\circ$).
The functions $a(t)$ and $b(t)$ both depend on time and location
of source and detector, but are independent of the polarization angle
$\psi$.

In the detector frame the phase of a gravitational wave produced by an
isolated neutron star can be written as~\cite{Jaranowski:1998qm}
\begin{equation}
\Psi(t) = \Phi_{0} + 2\pi\displaystyle\sum_{k=0}^{s} f_{0}^{(k)}
\frac{t^{k+1}}{(k+1)!} 
+ \frac{2\pi}{c}\textbf{n}_{0}\cdot\textbf{r}_{d}(t)
\displaystyle\sum_{k=0}^{s}f_{0}^{(k)}\frac{t^{k}}{k!},
\label{eq4}
\end{equation}
where $\Phi_{0}$ is the phase at the start time of the observation,
$f_{0}^{(k)}$ is the $k^\mathrm{th}$ derivative of the frequency, $c$ is the
speed of light, $\alpha$ and $\delta$ are the right ascension and
declination of the source, $\textbf{n}_{0}=\textbf{n}_{0}(\alpha,\delta)$ is the
unit vector of the source in the Solar System barycenter (SSB)
reference frame, $\textbf{r}_{d}$ is the position vector of
the detector in the same frame, and $s$ is the order of the expansion.
Neglecting changes in the proper motion of the star, the third term in Eq.~(\ref{eq4}) is a correction to the phase due to
the detector motion relative to the neutron star.

We can define $\Phi(t) = \Psi(t) - \Phi_{0}(t)$, as well as defining
\begin{equation}
\Phi_s(t) = 2\pi\displaystyle\sum_{k=1}^{s} f_{0}^{(k)}
\frac{t^{k+1}}{(k+1)!} 
+ \frac{2\pi}{c}\textbf{n}_{0}\cdot\textbf{r}_{d}(t)
\displaystyle\sum_{k=1}^{s}f_{0}^{(k)}\frac{t^{k}}{k!}
\label{eq4pt1}
\end{equation}
and
\begin{equation}
t_{m} = \frac{\textbf{n}_{0}\cdot\textbf{r}_{d}(t)}{c}.
\label{eq4pt2}
\end{equation}
Equations~(\ref{eq4pt1}) and (\ref{eq4pt2}) let us write

\begin{equation}
\Phi(t) = 2\pi f[t+t_{m}(t;\alpha,\delta)]+\Phi_{s}(t;f^{(k)},\alpha,\delta),
\label{eq5}
\end{equation}
which has the modulation due to the detector's motion around the SSB
clearly separated from the modulation due to the gravitational wave's
intrinsic frequency, although not the derivatives of the frequency.

An almost optimal statistic for the detection of continuous gravitational
wave signals is called the \F-statistic~\cite{Jaranowski:1998qm,Prix:CQG}. It
is the logarithm of the likelihood function maximized over the
intrinsic and unknown signal parameters. The \F-statistic is given by
\begin{equation}
\mathcal{F} = \frac{4}{S_{h}(f)T_{0}}\frac{B|F_{a}|^{2}+A|F_{b}|^{2}-2C\textit{R}(F_{a}F_{b}^{*})}{D}.
\label{eq6}
\end{equation}
where $S_{h}(f)$ is the one-sided spectral density of the detector's
noise at frequency $f$ and $T_{0}$ is the observation time.  $A$,
$B$, $C$, and $D$ are given by
\begin{equation}
A = (a \| a);
B = (b \| b);
C = (a \| b);
D = A\cdot B - C^{2}
\label{eq9}
\end{equation}
with
\begin{equation}
(x \| y) = \frac{2}{T_{0}}\displaystyle
\int^{\frac{T_{0}}{2}}_{\frac{-T_{0}}{2}}x(t)y(t) dt.
\label{eq10}
\end{equation}
$F_{a}$ and $F_{b}$ are integrals defined as
\begin{equation}
F_{a}(f) =
\displaystyle\int_{\frac{-T_{0}}{2}}^{\frac{T_{0}}{2}}a(t)x(t)
e^{-\textit{i}\Phi(t)} dt
\label{eq7}
\end{equation}
and
\begin{equation}
F_{b}(f) =
\displaystyle\int_{\frac{-T_{0}}{2}}^{\frac{T_{0}}{2}}b(t)x(t)
e^{-\textit{i}\Phi(t)} dt.
\label{eq8}
\end{equation}

We define a new time variable called $t_{b}$ as follows:
\begin{equation}
t_{b} = t + t_{m}.
\label{eq11}
\end{equation}

Taking a derivative with respect to $t$ on both sides of Eq.~(\ref{eq11}), we get

\begin{equation}
\frac{dt_{b}}{dt} = 1 + \frac{dt_{m}}{dt}
\label{eq11b} 
\end{equation}

From Eqs.~(\ref{eq4pt2}) and (\ref{eq11b}), we get

\begin{equation}
\frac{dt_{m}}{dt} = \frac{\textbf{n}_{0}\cdot\textbf{v}_{d}(t)}{c}
\label{eq11c}
\end{equation}

where $\textbf{v}_{d}(t)$ is the velocity of the detector in the SSB frame and thus $ \frac{\textbf{n}_{0}\cdot\textbf{v}_{d}(t)}{c} $ is the Doppler shift of the source with respect to the detector. For a detector located on Earth, the maximum Doppler shift experienced is of the order of $10^{-4}$. Using this fact and equation~\ref{eq11b} we get $\delta t_{b} \approx \delta t$.

We can thus rewrite the Eqs. for $F_{a}$ and $F_{b}$ as
\begin{equation}
F_{a}(f) =
\displaystyle\int_{\frac{-T_{0}}{2}}^{\frac{T_{0}}{2}}a(t_{b})x(t_{b})
e^{-2\pi\textit{i}ft_{b}}e^{\textit{i}\Phi_{s}(t_{b})} dt_{b},
\label{eq12}
\end{equation}
and
\begin{equation}
F_{b}(f) =
\displaystyle\int_{\frac{-T_{0}}{2}}^{\frac{T_{0}}{2}}b(t_{b})x(t_{b})
e^{-2\pi\textit{i}ft_{b}}e^{\textit{i}\Phi_{s}(t_{b})} dt_{b}
\label{eq13}
\end{equation}
which are just the Fourier transforms of the resampled data and the
detector response, multiplied by a phase
$e^{\textit{i}\Phi_{s}(t_{b})}$~\cite{Jaranowski:1998qm}.
Eqs.~(\ref{eq12}) and (\ref{eq13}) can be efficiently evaluated using FFTs.  Details of the resampling procedure can be
found in Sec.~\ref{resamp}.

\section{Implementation of barycentric resampling}
\label{sec:implement_bary_resamp}

Gravitational wave detectors collect data at the rate of about 16-20
kHz for spans of time on the order of a year. This means that typical
searches for gravitational waves will involve on the order of a terabyte (TB)
of data.  Computers currently have memories of a few
gigabytes (GB), making it necessary to break up the data into pieces
that can fit in the memory of a single computer.  To analyze the
full data set hundreds to thousands of these computers can then be
used together in the form of a Beowulf cluster, or tens to hundreds of
thousands with distributed computing systems such as
Einstein@Home~\cite{Abbott:2008uq}.

\subsection{Time Domain Analysis}

The \F-statistic can be calculated from a time series directly by
following the steps outlined in Section~\ref{sec:signal}.  However,
due to the large amounts of data involved, it is impractical to do
this for the entire data set.  One way to address this problem is to divide the data into band-limited time series, making
it possible to analyze one small sub-band at a time. Time series
spanning different frequency bands are then analyzed in parallel
on a Beowulf cluster or a distributed computing system.  In this
section we provide details on how this is accomplished in the time
domain, and address some of the difficulties that arise.

\subsubsection{Heterodyning, low-pass filtering, and downsampling}
\label{TimeHetLowDown}

Let the output of the instrument be the time series $x(t)$, and its 
Fourier transform be
\begin{equation}
\tilde{x}(f)=\displaystyle\int_{-\infty}^{\infty}x(t) e^{-2\pi\textit{i}ft} dt.
\label{eq14}
\end{equation}
If we consider the Fourier transform of the complex time series 
$x_h(t)=x(t)e^{2\pi \textit{i}f_ht}$,
\begin{eqnarray}
\tilde x_h(f)&=&\displaystyle\int_{-\infty}^{\infty}x(t)e^{2\pi \textit{i}f_ht}e^{-2\pi \textit{i}ft}dt 
\nonumber
\\
&=& \displaystyle
\int_{-\infty}^{\infty}x(t)\cdot e^{-2\pi \textit{i} (f-f_h)t}dt 
\nonumber
\\
&=&\displaystyle
\tilde{x}(f-f_h),
\label{eq15}
\end{eqnarray}
it is obvious that multiplying the time series $x(t)$ by $e^{2\pi \textit{i}f_ht}$
has shifted all the frequencies in the time series $x(t)$ by $f_h$. 
This procedure is referred to as complex heterodyning. 

If just a small frequency band $B$ of data around $f_h$ is of interest,
low-pass filtering followed by downsampling can be used to reduce the
bandwidth of the data appropriately. Specifically, if we wish to downsample by a
factor $D$, the new Nyquist frequency of our time series will be given
by
\begin{equation}
f_{{\rm Nyq,new}} = \frac{f_{{\rm Nyq,old}}}{D}=\frac{B}{2}.
\label{eq16}
\end{equation}
A simple but effective downsampling technique involves picking every
$D^\mathrm{th}$ point in the time series. To avoid aliasing effects however,
prior to downsampling a low pass filter must be applied to the data
with a sharp fall-off around the new Nyquist frequency.  The
heterodyned, band-limited, downsampled complex time series will have a
sampling time $\Delta t = {1}/{B}$. For example, suppose we are only interested in analyzing data between 990 Hz and 1 kHz. By multiplying the data with the phase factor $e^{2\pi (995) \textit{i} t}$, data at 995 Hz moves to 0 Hz (DC), 990 Hz moves to -5 Hz, and 1 kHz to +5 Hz (we have taken $t$ to be measured in seconds). To avoid aliasing problems when we downsample, we low-pass filter the data at 5Hz, the new Nyquist frequency. We can then downsample by picking one point out of every 100.  The resulting complex time series will be sampled at 10 Hz and contain all the information in the original time series between 990 Hz and 1 kHz.

\subsubsection{Barycentric resampling and heterodyne correction}
\label{resamp}

In this section we explain how to use the low bandwidth heterodyned
complex time series to compute the \F-statistic given by
Eq.~(\ref{eq6}).

In the following we will work only with $F_{a}$. The procedure for
$F_{b}$ is completely analogous. It is easiest to begin with the
integral definition for $F_{a}$ in Eq.~(\ref{eq7}) with the phase
explicitly written out, namely,
\begin{equation}
F_{a}(f) =
\displaystyle\int_{\frac{-T_{0}}{2}}^{\frac{T_{0}}{2}}a(t)x(t)
e^{-2\pi\textit{i}f (t + t_{m}) }e^{\textit{i}\Phi_{s}(t)} dt,
\label{Fa1}
\end{equation}
and a similar expression holds for $F_{b}$.  The heterodyned version
of $F_{a}$ is
\begin{equation}
F_{a}(f-f_h) =
\displaystyle\int_{\frac{-T_{0}}{2}}^{\frac{T_{0}}{2}}a(t)x(t)
e^{-2\pi\textit{i}(f-f_h) (t + t_{m}) }e^{\textit{i}\Phi_{s}(t)} dt.
\label{Fa2}
\end{equation}
If we already have a complex heterodyned 
time series $x_h(t)$ (heterodyned in the detector frame), we can
use it to absorb some (but not all) of the heterodyne exponent in
Eq.~(\ref{Fa2}) as follows:
\begin{eqnarray}
x(t) e^{-2\pi\textit{i}(f-f_h) (t + t_{m}) } = 
x_h(t) e^{2\pi\textit{i} f_h t_{m}} e^{-2\pi \textit{i} f
  (t + t_{m})}.
\nonumber
\\
\label{absorb}
\end{eqnarray}
This means that rather than  Eq.~(\ref{Fa2}), we should evaluate
\begin{equation}
F_{a}(f-f_h) =
\displaystyle\int_{\frac{-T_{0}}{2}}^{\frac{T_{0}}{2}}a(t)z(t)
e^{-2\pi\textit{i}f (t + t_{m}) }e^{\textit{i}\Phi_{s}(t)} dt,
\label{Fa3}
\end{equation}
where
\begin{equation}
z(t)=x_h(t) e^{2\pi\textit{i} f_h t_{m}}.
\end{equation}
At this point we have an expression which looks like Eqs.~(\ref{eq7})
and (\ref{eq8}), and we can write the integral over $t$ instead as 
an integral over $t_b$: 
\begin{equation}
F_{a}(f-f_h) =
\displaystyle\int_{\frac{-T_{0}}{2}}^{\frac{T_{0}}{2}}a(t_{b})z(t_{b})
e^{-2\pi\textit{i}ft_{b}}e^{\textit{i}\Phi_{s}(t_{b})} dt_{b},
\label{Fa4}
\end{equation}
with a similar expression for $F_{b}$.

The discrete version of Eq.~(\ref{Fa4}) for a time series with $N$ points reads
\begin{equation}
F_{a}(f-f_h) = \displaystyle\sum_{k = 1}^{N}a(t_{b}^{k})z(t_{b}^{k})
e^{-2\pi\textit{i}ft_{b}^{k}}e^{\textit{i}\Phi_{s}(t_{b}^{k})}dt_{b},
\label{eq19}
\end{equation}
and a similar expression holds for $F_{b}$:
\begin{equation}
F_{b}(f-f_h) = \displaystyle\sum_{k = 1}^{N}b(t_{b}^{k})z(t_{b}^{k})
e^{-2\pi\textit{i}ft_{b}^{k}}e^{\textit{i}\Phi_{s}(t_{b}^{k})}dt_{b},
\label{eq20}
\end{equation}
where $t_{b}^{k}$ is the $k^\mathrm{th}$ datum in the time series as measured
in the barycentric frame and $dt_{b} = t_{b}^{k+1}-t_{b}^{k}$.  The 
relationship between $t_{b}$ and $t$ can be written as
\begin{equation}
t_{b}^{k} = t^{k}+t_{m}(t^{k};\alpha,\delta).
\label{eq21}
\end{equation}
This relationship between $t^{k}$ and $t_{b}^{k}$ can be used to
calculate $z(t_{b}^{k})$ from the time series $z(t^{k})$. In
practice, one starts out with $z(t^{k})$, i.e. data sampled
regularly in the detector frame. Then we calculate
${T}^{k}(t_{b}^{k})$, which are detector times corresponding to
regularly spaced samples in the barycentric frame. These
$T^{k}(t_{b}^{k})$ are irregularly sampled in the detector frame,
but since we have $z(t^{k})$, we can calculate
$z(T^{k}(t_{b}^{k}))$ by using interpolation. The interpolated
time series $z(T^{k}(t_{b}^{k}))$ is the $z(t_{b}^{k})$ of
Eqs.~(\ref{eq19}) and (\ref{eq20}).  A similar procedure may be used
to calculate the $a(t_{b}^{k})$ from $a(t^{k})$, and the
$b(t_{b}^{k})$ from $b(t^{k})$. The factor of
$e^{\textit{i}\Phi_{s}(t_{b}^{k})}$ in Eqs.~(\ref{eq19}) and
(\ref{eq20}) is calculated using equation ~(\ref{eq4pt1}). In this
case, instead of calculating $\Phi_{s}(t^{k})$, we calculate
$\Phi_{s}(T^{k}(t_{b}^{k}))$, which is equivalent to calculating
$\Phi_{s}(t_{b}^{k})$. While in theory one has to calculate the
quantity $\textbf{n}_{0}\cdot\textbf{r}_{d}(t)$ in equation
~(\ref{eq4pt1}), in practice this information is already encoded in
$T^{k}(t_{b}^{k})$ as
\begin{equation}
\label{phi-s-explain}
\textbf{n}_{0}\cdot\textbf{r}_{d}(t) = t_{m}\cdot c 
= (t_{b}^{k}-T^{k}(t_{b}^{k}))\cdot c \;.
\end{equation}
With all the parts of Eqs.~(\ref{eq19}) and (\ref{eq20}) in hand, we
can compute $F_{a}(f-f_h)$ and $F_{b}(f-f_h)$.

\begin{figure}[t]
\begin{center}
\includegraphics[width=3.5in]{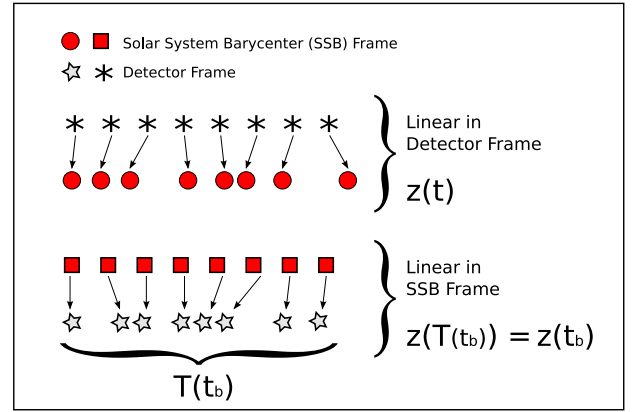}
\caption{Graphical description of the resampling procedure}
\label{fig3}
\end{center}
\end{figure}

In summary, the procedure is the following:
\begin{enumerate}
\item Start with a heterodyned, band-limited, downsampled
  $x_h(t^{k})$ with $t^{k}$ regularly spaced in time, in 
the frame of reference of the detector.
\item Correct the $x_h(t^{k})$ for the heterodyning done in the detector frame 
by multiplying with $e^{2\pi\textit{i} f_{h} t_{m}}$ to produce the
$z(t^{k})$.
\item The $z(t^{k})$ correspond to data irregularly
spaced in the barycentric frame. Calculate $T^{k}(t_{b}^{k})$,
which are times in the detector frame corresponding to regularly
sampled solar system barycenter times.
\item Using interpolation, calculate $z(T^{k}(t_{b}^{k}))$ from $z(t^{k})$,
which is the $z(t_{b}^{k})$ used in Eqs.~(\ref{eq19}) and (\ref{eq20}).
\item Similarly, from $a(t^{k})$ and $b(t^{k})$ calculate 
$a(t_{b}^{k})$ and $b(t_{b}^{k})$ respectively. 
\item Using FFTs, evaluate Eqs.~(\ref{eq19}) and (\ref{eq20}) 
to calculate $F_{a}(f-f_h)$ and $F_{b}(f-f_h)$. 
\item Use Eq.~(\ref{eq6}) to calculate the \F-statistic.
\end{enumerate}

\subsection{Frequency Domain Analysis} 

In the previous section we describe a practical way of calculating the
\F-statistic from time series data. However, in practice the
calculation is done in the frequency domain for a couple of reasons.
One is that much of the code written in the LIGO
Scientific Collaboration's (LSC) Continuous Waves working group is
tailored to an analysis performed in the frequency domain and hence
there exist many data processing and validation tools to process the data that are useful to this code. Another reason is that
gravitational wave detectors are subject to many sources of noise,
some of which change daily or even hourly, such as wind, microseism,
earthquakes, anthropogenic noise, etc. These change the noise floor of
any analysis as a function of time. Working in the frequency domain is
a natural way to deal with this problem.

We begin a frequency domain analysis by taking short time-baseline
Fourier transforms of the time domain data, called short Fourier
transforms (SFTs). When we calculate the \F-statistic, we divide 
by the noise in the instrument at that frequency, as shown in 
Eq.~(\ref{eq6}).  However, Eq.~(\ref{eq6}) assumes the noise is 
stationary.  To account for the non-stationarity of the noise we need 
to weight by the noise over time, which is done on a per SFT basis. This 
normalization process is described in the next section.

The computational cost of estimating the noise per SFT scales with the
number of SFTs and thus for a fixed observation time scales
inversely with the time-baseline. A compromise is needed between the
demands of computational time and relative stationarity of the
detector for a given time-baseline. In LIGO, SFTs are usually 1800
seconds long, since the detector is reasonably stationary for that
time.

\subsubsection{Dealing with non-stationary and colored data}
\label{NonstationaryNoise}
To deal with non-stationarities, variations in the noise floor from
SFT to SFT, and colored data, we can normalize our SFT data to absorb
the $1/S_h(f)$ term in the definition of the \F-statistic in
Eq.~(\ref{eq6}). If $X_{\alpha, k}$ is the $k^\mathrm{th}$ frequency bin 
of the $\alpha^\mathrm{th}$ SFT, then we can redefine a normalized data 
point $\hat X_{\alpha, k}$ as
\begin{equation}
X_{\alpha, k} \longrightarrow \hat X_{\alpha, k}=\frac{X_{\alpha, k}}{\sqrt{S_{\alpha,k}}}\;,
\label{dimlessSFTs1}
\end{equation}
where $S_{\alpha,k}$ is an estimate of the one-sided power spectral
density for the $k^\mathrm{th}$ frequency bin of the $\alpha^\mathrm{th}$ SFT.
Estimators used for this purpose should be robust in the presence of
spectral features in the data, such as a running median.

\subsubsection{Merging SFTs into long time-baseline Fourier transforms}

There are many practical difficulties that arise when dealing with
SFTs. Often contiguous chunks of data have to be divided up into
multiple SFTs and it is necessary to coherently combine them into one
long time-baseline SFT. This is done using the Dirichlet kernel, which
is the equivalent of a sinc interpolation (ideal interpolation) done
in the time domain. In order to keep the computational cost down, the
Dirichlet kernel is truncated at a finite number of points (usually
around 16). This introduces a slight interpolation error, which cannot
be avoided without sacrificing a large amount of computational power.

Suppose we divide the data $x(t)$ of length $T_0$ into $M$ short chunks of 
length $T_{\rm SFT}$ each with $N$ points, so that $T_0 = MT_{\rm SFT}$. The
discrete Fourier transform (DFT) of the data is
\begin{eqnarray}
\label{e1}
X_{b}=\sum_{l=0}^{\it{NM-1}} x_{l} e^{-2{\pi}i lb/NM},
\end{eqnarray}
where $x_{l}=x(l\Delta t)$, $\Delta t$ is the sampling time, and $b$
is a long time-baseline frequency index. We
can write the Fourier transform in terms of two sums:
\begin{eqnarray}
\label{dft1}
X_{b}=\sum_{\alpha=0}^{\it{M-1}} \sum_{j=0}^{\it{N-1}} x_{\alpha,j}
e^{-2{\pi}ib(j+N\alpha)/NM}\;, 
\end{eqnarray}
where $x_{\alpha,j}=x((j+N\alpha)\Delta t)$. We can express the
$x_{\alpha,j}$ in terms of an inverse DFT of a short chunk of data,
\begin{equation}
x_{\alpha,j}=\frac{1}{N}\sum_{k=0}^{N-1}X_{\alpha,k}\, e^{2\pi{i}jk/{N}},
\label{e3}
\end{equation}
where the $X_{\alpha,k}$ are the starting SFT data, 
\begin{equation}
X_{\alpha,k}=\sum_{j=0}^{N-1}x_{\alpha,j}\, e^{-2\pi{i}jk/{N}}.
\label{e25}
\end{equation}
Replacing $x_{\alpha,j}$ with Eq.~(\ref{e3}) in Eq.~(\ref{dft1}) gives
\begin{eqnarray}
\label{dft2}
X_{b}&=&\sum_{\alpha=0}^{\it{M-1}} \sum_{j=0}^{\it{N-1}} \left(
  \frac{1}{N}\sum_{k=0}^{N-1}X_{\alpha,k}\, e^{2\pi{i}jk/{N}} \right)
e^{-2{\pi}ib(j+N\alpha)/NM} 
\nonumber
\\
&=&
\frac{1}{N}\sum_{\alpha=0}^{\it{M-1}} e^{-2\pi i b \alpha/M} 
\sum_{k=0}^{N-1}X_{\alpha,k}
\sum_{j=0}^{\it{N-1}} 
e^{-2{\pi}i j (b/M-k)/N}\;.
\nonumber
\\
\end{eqnarray}
The last sum in this expression can be evaluated analytically. In particular,
\begin{equation}
\sum_{j=0}^{\it{N-1}} z^{cj}=\frac{1-z^{Nc}}{1-z^c}.
\end{equation}
We take $z=e$, $c=-iy/N$, with $y=2\pi(b/M- k)$, so that the sum is given by
\begin{equation}
\sum_{j=0}^{\it{N-1}} e^{-iyj/N}=
\frac{1-e^{-iy}}{1-e^{-iy/N}}\;.
\end{equation}
In the large $N$ limit the exponent of the denominator will be small so that
\begin{eqnarray}
\frac{1-e^{-iy}}{1-e^{-iy/N}} &\approx& \frac{1-e^{-iy}}{1-(1-iy/N)}
=\frac{iN}{y}(e^{-iy}-1)
\nonumber
\\
&=&N(\frac{\sin y}{y}-i \frac{1-\cos y}{y}).
\label{e75}
\end{eqnarray}
This means we can write Eq.~(\ref{dft2}) as 
\begin{eqnarray}
\label{dft3}
X_{b} =\sum_{\alpha=0}^{\it{M-1}} e^{-2\pi i b \alpha/M} 
\sum_{k=0}^{N-1}X_{\alpha,k} P_{b,k},
\end{eqnarray}
with the Dirichlet kernel
\begin{equation}
P_{b,k}=\frac{\sin y}{y}-i \frac{1-\cos y}{y},
\end{equation}
and $y=2\pi(b/M- k)$.
The function $P_{b,k}$ is very strongly peaked around $y=0$, which
is near a value of the frequency index $k^*={\rm floor}(b/M)$. 
This means one only needs to evaluate the sum over $k$ for a few terms $\Delta k$ around $k^*$. With this in
mind we write
\begin{equation}
\label{e9}
X_{b} \approx
\sum_{\alpha=0}^{\it{M-1}} e^{-2\pi i b \alpha/M} 
\sum_{k=k^*-\Delta k}^{k^*+\Delta k} X_{\alpha,k}P_{\alpha,k}.
\end{equation}
To produce a heterodyned time series a sub-band of the $X_{b}$
may be selected and inverse Fourier transformed.

\subsubsection{Normalized long time-baseline Fourier transforms}
With the normalized SFT data $\hat X_{\alpha, k}$ from
Eq.~(\ref{dimlessSFTs1}) 
we can construct a normalized version of the long time-baseline 
Fourier transform
\begin{equation}
\label{e91}
\hat X_{b} \approx
 \sum_{\alpha=0}^{\it{M-1}} e^{-2\pi i b \alpha/M} 
 \sum_{k=k^*-\Delta k}^{k^*+\Delta k} \hat X_{\alpha,k}P_{\alpha,k},
\end{equation}
and take a sub-band of $\hat X_{b}$, inverse Fourier
transform it, and produce the heterodyned time series, and correct it to produce $\hat
z(t_b^k)$.  In terms of this time series, we can write
\begin{align}
\hat F_{a}(f-f_h) = \displaystyle\sum_{k = 1}^{N}\hat z(t_{b}^{k})a(t_{b}^{k})
e^{-2\pi\textit{i}ft_{b}^{k}}e^{\textit{i}\Phi_{s}(t_{b}^{k})}
\label{Fatilde}
\end{align}
and
\begin{align}
\hat F_{b}(f-f_h) = \displaystyle\sum_{k = 1}^{N}\hat z(t_{b}^{k})b(t_{b}^{k})
e^{-2\pi\textit{i}ft_{b}^{k}}e^{\textit{i}\Phi_{s}(t_{b}^{k})}\;,
\label{Fbtilde}
\end{align}
and thus
\begin{equation}
\mathcal{F} = \frac{4}{T_0} \frac{B|\hat{F}_{a}|^{2}
+A|\hat{F}_{b}|^{2} - 2C \Re(\hat{F}_{a}\hat{F}_{b}^{*})}{D}.
\label{LALDemodFstat}
\end{equation}

\subsubsection{Heterodyning}
As shown before in Eqs.~(\ref{eq14}) and (\ref{eq15}), heterodyning
is a procedure by which the frequency of interest can be shifted
arbitrarily. When one applies the kind of correction in
Eq.~(\ref{eq14}), we effectively move all the frequencies by a set amount.
By doing so, we convert the time series from a real time series to
a complex time series, with the same amount of information content.

Heterodyning in the frequency domain can be done in two ways, one in
which the time series produced after inverse Fourier transforming is
real and another in which it is complex.  A cosine transform used to
heterodyne would produce a real time series, but this method is not
used in an implementation of the technique (see section~\ref{sec:results}).  A complex heterodyned time series is produced by inverse Fourier transforming a 
relabelled band of the frequencies.  Since in Eq.~(\ref{eq14}), all
frequencies are shifted by a fixed amount, the equivalent procedure in
the frequency domain is just relabelling the heterodyne frequency
$f_{h}$ as DC and subsequently all the other frequencies relative to
this new DC.

Taking the example from Section~\ref{TimeHetLowDown}, we can just
internally change the labels of the 995 Hz frequency bin to DC
and 1000 Hz to 5 Hz. Once this relabelling is done, the original
data will have all shifted by 995 Hz, with the 10 Hz from -5 Hz to
+5 Hz containing all the relevant information.  If one were using the
whole band without downsampling or filtering, then this relabelling
would have to wrap around the Nyquist frequency edge, but since the
whole purpose of heterodyning is to downsample, it is never necessary
to do so.

\subsubsection{Downsampling and low-pass filtering}
Following the time domain algorithm, after heterodyning the data, it
needs to be downsampled and low-pass filtered. The downsampling and low-pass filtering is achieved by simply throwing out the data that is not in the band of
interest. The heterodyning is done in such a way as to keep the center
of the band of interest at DC. A Tukey window applied to the band of interest, keeping a little bit of
data on both edges to facilitate the rise of the window
from 0 to 1, is a good choice of a low-pass filter. Once an inverse Fourier transform is performed on this
smaller subset in the frequency domain, it generates the same
heterodyned, downsampled, and low-pass filtered time series as the
time domain algorithm.

\subsubsection{Gaps in the data}
Data collected by an interferometer will have gaps due to periods of
downtime. These gaps need to be dealt
with in a manner that preserves the phase coherence of the segments
around the gaps. The gaps increase the analysis
time without contributing any power to the \F-statistic, and thus act
like a zero padding.

The data is divided up into a series of contiguous chunks and gaps.
For each contiguous chunk the SFTs in that chunk are normalized,
patched up and then a heterodyned, downsampled and low-pass-filtered
time series is calculated from it.  Heterodyning done by relabelling
is equivalent to multiplying with $e^{2\pi\textit{i}f_h(t-t_c)}$,
where $t_c$ is the start time of the data chunk being heterodyned and
$f_h$ is the heterodyne frequency.  If we have multiple chunks that
are separately being heterodyned, then $t_c$ is different for each
chunk.  In the time domain analysis, we assumed that the heterodyne
reference time is the same as the start time of the analysis.  In
order to achieve the same kind of heterodyning, one needs to multiply
each newly created time series with a correcting phase factor, namely
\begin{equation}
e^{2\pi\textit{i}f_h(t_c - t_s)},
\label{hetStartCorr}
\end{equation}
where $t_s$ is the start time of the overall analysis.

A Tukey window can then be applied to each of these time series to
smoothly bring the data to zero at the edges, which correspond to the
gaps. The gaps are then filled with zeros, as no data was collected
during those times. This procedure is repeated for all the gaps and
contiguous chunks. At the end, a time series is produced, which is
contiguous and spans the time of the analysis.  By ensuring that the
timestamps of the first datum of each contiguous chunk correspond
with the start time of that chunk, we ensure that the phase coherence
is maintained throughtout.

\subsubsection[*]{Summary}
To summarize, a simple algorithm to produce a time series equivalent to the
one used for the time domain analysis is as follows:
\begin{enumerate}
\item Divide the data into time chunks and Fourier transform them to create SFTs.
\item Normalize these SFTs and assign them weights.
\item Identify contiguous sets of SFTs.
\item Combine each contiguous chunk of SFTs into one long time-baseline Fourier transform (FT).
\item Create a downsampled, heterodyned, and low-pass-filtered time series by inverse Fourier transforming the desired frequencies from the FT.
\item Stitch all these time domain chunks together, filling gaps with zeros.
\end{enumerate} 

\begin{figure}[t]
\begin{center}
\includegraphics[width=3in]{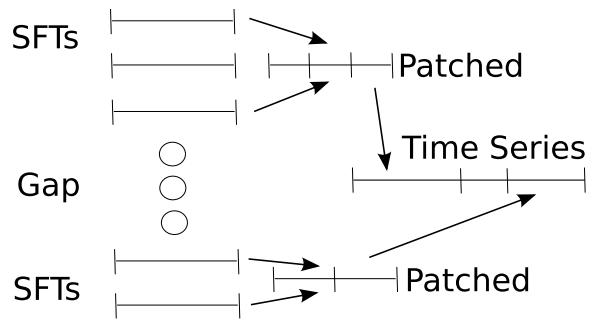}
\caption{Pictorial description of data pre-processing}
\label{fig1}
\end{center}
\end{figure}

\section{Results}
\label{sec:results}
\subsection{Speed}
The scheme previously used to compute the \F-statistic, involved the use of the Dirichlet kernel to combine a series of SFTs ~\cite{V2Document} ~\cite{Williams:1999}, which were calculated for 30 minutes of data taken at 16 kHz. The 30 minute window was set by the maximum Doppler shift due to the motion of the Earth. A C code called ComputeFStatistic$\_$v2~\cite{LAL:V2} was written in the LIGO Analysis Library (LAL) to calculate the \F-statistic using this algorithm. The code which implements our method is also written in C and is called ComputeFStatistic$\_$resamp~\cite{LAL:Resamp}. Henceforth we will refer to the previous implementation as the LAL implementation and our implementation as Resampling. 

The \F-statistic is calculated for a series of templates looping over various parameters such as sky location, $\alpha$ and $\delta$, spin-downs $f^{k}$, and various frequencies $f$. We can ignore the way the two implementations deal with loops over $\alpha$, $\delta$, and $f^{k}$, since they both loop over them in the same manner. The speed of computation for a loop over frequencies $f$ is worth comparing, however.  

Assume that we have N data points (take for example $10^{6}$ seconds of data at
100 Hz, i.e. $10^{8}$ data points). Now assume that the number of operations per sky location and per spin-down is $N_{\mathrm{ops}}$. If the number of Dirichlet kernel points used is $N_{\mathrm{Dir\_Ker}}$, then the total number of operations used by
the LAL implementation is:

\begin{equation}
N^{\mathrm {LAL}}_{\mathrm{Tot}} = N_{\mathrm{ops}}\cdot N_{\mathrm{Dir\_ Ker}}\cdot N_{\mathrm{SFTs}}\cdot N\;,
\end{equation}
Where $N_{\mathrm{ops}}$ is defined as the number of operations conducted
in the innermost loop and is approximately of order 10, $N_{\mathrm
  {Dir\_Ker}}$ is the number of times the Dirichlet Kernel loop is
repeated, $N_{\mathrm {SFTs}} = \frac{T_{obs}}{\mathrm{T_{SFT}}}$ is the number
of SFTs, and $N$ is the number of data points.

Compare this to the resampling method, which consists of 4 major steps:

\begin{enumerate}
\item Calculating $t_{b}(t)$, given a sky location and time.
\item Calculating the integrands of $F_{a}$ and $F_{b}$.
\item Interpolating and calculating the beam patterns. 
\item Taking the Fourier transform.
\end{enumerate}

Each of these steps involves order 10 operations, but all of these
steps are sequential, therefore they only add, resulting in a total number of operations per data point,$N_{\mathrm {ops}}^{\mathrm {Resamp}}$, of approximately 30 operations. The last step is the Fourier
transform, which is of order $N\log{N}$, therefore the total number of
steps is:
\begin{equation}
N^{\mathrm{Resamp}}_{\mathrm {Tot}} = (N_{\mathrm {ops}}^{\mathrm {Resamp}}+\log{N})\cdot N \;.
\end{equation}
Therefore the ratio of operations between the two methods is 
\begin{equation}
\frac{N^{\mathrm{LAL}}_{\mathrm{Tot}}}{N^{\mathrm{Resamp}}_{\mathrm{Tot}}} =
\frac{N_{\mathrm{ops}}
\cdot N_{\mathrm{Dir\_ Ker}}\cdot N_{\mathrm {SFTs}}}{N_{\mathrm {ops}}^{\mathrm {Resamp}}+\log{N}}
\end{equation}
To first order, we have 
\begin{equation}
\frac{N^{\mathrm {LAL}}_{\mathrm{Tot}}}{N^{\mathrm{Resamp}}_{\mathrm{Tot}}} \approx \frac{N_{\mathrm{SFTs}}}{\log{N}}\;.
\end{equation}    

Therefore for large observation times, this method of calculating the
\F-Statistic is faster and, in the case of a targeted search, it allows
for a large parameter space in $F^{(k)}$'s. 

The speed-up in practice is reduced by a few practical issues as seen in section~\ref{sec:practical}. However, Resampling is still considerably more efficient than the LAL implementation. For Einstein@Home, because of the relatively small coherent integration time, the speed-up is around $10$. But for targeted searches that span multiple months or years, the improvement can be as high as a factor of $2000$. Thus, while some targeted searches which integrate over a couple of years were impossible to do previously, they are now possible.

\subsection{Validations}

The probability density distribution of the \F-statistic for Gaussian
noise of zero mean and unity standard deviation is a $\chi^{2}$
distribution with four degrees of freedom. In the presence of a
signal, the distribution is a $\chi^{2}$ of four degrees of freedom
with a non-centrality parameter given by the \F-statistic in the
absence of noise for the particular signal.

\begin{figure}[t]
\begin{center}
\includegraphics[width=3.5in]{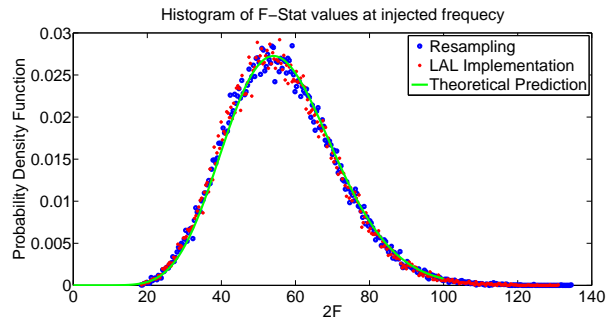}
\caption{Histogram of results of Monte Carlo simulation with signals injected in different instances of noise}
\label{fig2}
\end{center}
\end{figure}

Resampling uses various approximate methods in the calculation of the
\F-statistic, and this can lead to disagreements between the
theoretical \F-statistic probability density function and the output
of the code. These changes are of the order of a few percent and are
within acceptable limits. The validity of the code can be tested by
using a Monte Carlo simulation of about a million different injections
of the same signal in different instances of noise. The noise is
generated as a Gaussian noise of zero mean and unity standard
deviation, and the signal is added into this noise. For each individual injection the signal is
chosen with a given set of amplitude parameters and a fixed sky
location and spindowns, and the search is conducted over these exact chosen parameters in order to avoid any mismatches. These Monte Carlos are then repeated with another set of parameters, which are themselves chosen randomly.
While it is not an exhaustive test, randomly chosen parameters ensure
that we are not biased in the validation test.  The plot in figure
\ref{fig2} is produced by performing one such Monte Carlo simulation.
In this case, both the LAL implementation and Resampling were run
on the same set of data. The \F-statistic was picked out at the
appropriate frequency and this was repeated about a million times. A
histogram of these \F-statistic values was then plotted. As one can
see, there is very good agreement in between the expected distribution
of the \F-statistic and the two implementations.

\section{Practical Considerations}
\label{sec:practical}
\subsection{Discreteness}

In the implementation of the algorithm explained above, one major obstacle is the fact that the data collected by any physical instrument is discrete and thus must be handled appropriately. Take, for example, the heterodyne frequency used in the calculation. This frequency cannot be chosen arbitrarily, as only certain frequencies are sampled and thus there are only certain permitted choices. 

Most major FFT computation algorithms output the frequency series in a specific format, which split the data into two parts. The first bin output by these algorithms is the DC followed by the first positive frequency bin up to positive Nyquist and then follows this up with the negative frequencies starting at the negative Nyquist frequency. This order of placing frequency bins speeds up computation and is necessary for the internal workings of these algorithms. Thus when an inverse FFT is performed on the frequency domain data in the form of SFTs, a simple reshuffling needs to be done. The frequency selected to be the first bin will become the new DC and thus the data will have been heterodyned by that said frequency. In order to ensure that the same frequency bin is chosen as DC, one needs an odd number of bins per SFT. If the number of bins are even, then upon increasing the amount of data it can shift this number to an odd number as the increase is always done by changing the number of SFTs. But if the number of bins per SFT is odd, then it will remain odd for any number of SFTs. This ensures that there is no mismatch in choosing the appropriate bin as the heterodyne frequency. 

\subsection{Interpolation Issue}
When the resampling algorithm is used on discrete data, one needs to
interpolate between data points to go from the detector frame to the
SSB frame. This interpolation acts like a nonlinear low-pass filter and destroys the power at higher frequencies in the band of analysis. Since the filter is nonlinear, the frequency response is not well-defined, and thus there is no way to compensate for the power loss at high frequencies. The power loss can be significant (of order $30\%$) and is unacceptable in most analyses.  The exact nature of the filter depends on the type of the interpolation routine used and the sky location that one resamples to. The only work around is to perform the computation over a larger band than the one desired. In practice it is sufficient to double the band and to discard the higher frequencies. 

\section{Summary and conclusions}

In this paper, we describe an efficient implementation of the
barycentric resampling technique, which deals with the
non-stationarity of the detector and calculates the \F-statistic.
Although the calculation of the \F-statistic has been targeted, this
technique can be used for many other kinds of searches. The major
contribution of this technique is to remove the Doppler shift of the
Earth's motion in a gravitational wave signal. Thus, once this Doppler
shift is removed, both frequentist and Bayesian techniques can be
applied to the data. In the process of implementing this algorithm, a series of practical issues are dealt with, including constraints of modern computer memory, discreteness of the data taken, losses due to interpolation, and gaps in real data.

The computational savings due to this technique can be used in various
ways. One such use is to increase the coherent integration time for
all-sky searches like the Einstein@Home searches. Currently
Einstein@Home~\cite{Abbott:2008uq} uses $40$ hour long coherent integration time. The
resampling code will be about $10$ times faster for such integration
times, and for the same computational power and keeping the same
scaling for the search, we can coherently integrate $64$ hours
instead, which corresponds to a sensitivity increase of about $25\%$.

The resampling technique is most effective for long integration times,
which are feasible for targeted searches like the search for
gravitational waves from the Crab pulsar~\cite{Abbott:2008cp}. The
computational savings can be used to search over wider parameter
spaces like more spindown parameters or to search over binary systems.

\acknowledgments

We would like to thank the membership of the LIGO Virgo Collaboration's continuous waves group, especially Stefano Braccini, Vladimir Dergachev, Greg Mendell, Chris Messenger, Marialessandra Papa and Reinhard Prix for helpful discussions. We are also grateful to Patrick Brady and Alan Weinstein for useful conversations. LIGO was constructed by the California Institute of Technology and Massachusetts Institute of Technology with funding from the National Science Foundation and operates under cooperative agreement PHY-0757058. This paper has LIGO Document Number LIGO-P0900301-v1. Xavier Siemens is supported in part by NSF Grant No. PHY-0758155 and the Research Growth Initiative at the University of Wisconsin-Milwaukee.

\end{document}